\documentstyle[12pt,graphicx,caption2]{article}

\begin{document}

\title{Field-theoretical Approach to Particle Oscillations in Absorbing Matter}
\author{V.I.Nazaruk\\
Institute for Nuclear Research of RAS, 60th October\\
Anniversary Prospect 7a, 117312 Moscow, Russia.*}

\date{}
\maketitle
\bigskip

\begin{abstract}
The $ab$ oscillations in absorbing matter are considered. The standard 
model based on optical potential does not describe the total $ab$
transition probability as well as the channel corresponding to absorption
of the $b$-particle. We calculate directly the off-diagonal matrix element
in the framework of field-theoretical approach. Contrary to one-particle
model, the final state absorption does not tend to suppress the channels
mentioned above or, similarly, calculation with hermitian Hamiltonian
leads to increase the corresponding values. The model reproduces all the
results on the particle oscillations, however it is oriented to the 
description of the above-mentioned channels. Also we touch on the problem 
of infrared singularities. The approach under study is infrared-free.
\end{abstract}

\vspace{5mm}
{\bf PACS:} 24.10.-i, 11.30.Fs, 13.75.Cs

\vspace{5mm}
Keywords: unitarity, self-energy, infrared singularity, diagram technique

\vspace{5cm}
*E-mail: nazaruk@inr.ru

\newpage

\setcounter{equation}{0}
\section{Introduction}
The theory of $ab$ oscillations is based on one-particle model [1-4]. The
interaction of particles $a$ and $b$ with the matter is described by a potentials
$U_{a,b}$. ${\rm Im}U_b$ is responsible for loss of $b$-particle intensity. The
wave functions $\Psi _{a,b}$ are given by equations of motion. The index of
refraction, the forward scattering amplitude $f(0)$ and potential are related to
each other, so later on the standard approach is referred to as potential model.

Description of absorption by means of ${\rm Im}U_b$ is at least imperfect and
should be partially revised. As an illustration, let us consider the case of strong
$b$-particle absorption. Instead of periodic process, we get two-step system decay
\begin{equation}
(a-\mbox{medium})\rightarrow (b-\mbox{medium})\rightarrow f.
\end{equation}
Here $(b-\mbox{medium})\rightarrow f$ represents the $b$-particle absorption. The
potential model does not describe this process as well as the total $ab$ transition
probability (see [5,6] and next section). It describes the probability of finding a
$b$-particle only.

By way of specific example we consider the $n\bar{n}$ transitions in medium
followed by annihilation [7-13]
\begin{equation}
(n-\mbox{medium})\rightarrow (\bar{n}-\mbox{medium})\rightarrow M,
\end{equation}
where $M$ are the annihilation mesons. The qualitative process picture is as follows.
The free-space $n\bar{n}$ transition comes from the exchange of Higgs bosons with the
mass $m_H>10^5$ GeV [8,9]. From the dynamical point of view this is a momentary
process: $\tau _c\sim 1/m_H<10^{-29}$ s. The antineutron annihilates in a time $\tau
_a\sim 1/\Gamma $, where $\Gamma $ is the annihilation width of $\bar{n}$ in the
medium. We deal with two-step process with the characteristic time $\tau _2\sim
\tau _a$.

The potential model describes the probability of finding an antineutron only, whereas
a main contribution gives the process (2) because $\bar{n}$ annihilates in a time
$\tau _a$.  In the following we consider the $n\bar{n}$ transitions since the final
state absorption in this case is extremely strong.

So the model should describe the two-step processes like (1) and (2). On the other
hand, as absorption Hamiltonian tends to zero, the well-known results on particle
oscillations should be reproduced. This program is realized below.

Also we present here an elaborated derivation of the lower limit on the $n\bar{n}$
oscillation time and discuss in some detail the uncertainties connected with
medium corrections.

The paper is organized as follows. In the next section, we recall the main results
and point out the chief drawback of the potential model. In this model the $n\bar
{n}$ transition probability depends critically on the antineutron self-energy
$\Sigma $. In the field-theoretical approach the similar picture takes place (Sects.
3 and 4). Because of this, the particular attention is given to the suppression
mechanism and origin of $\Sigma $. In Sect. 4 we arrive at the conclusion that
$\Sigma =0$. An important effect of the competition between scattering and
annihilation of $\bar{n}$ in the intermediate state is studied as well. The
main calculations are performed in Sects. 5 and 6. If $\Sigma =0$, the $S$-matrix
amplitudes contain the infrared singularity. For solving the problem the approach
with finite time interval (FTA) [14] is used. It is infrared free. First of all we
verify the FTA by the example of exactly solvable potential model. The FTA reproduces
all the results on the particle oscillations ($\nu _a\nu_b$, $n\bar{n}$, etc). In
Sect. 6 the process (2) and process shown in Fig. 8b are calculated. The linkage
between the $S$-matrix theory and FTA is studied as well. In Sect. 7 we complete the
calculation of process (2). Also we arrive at the conclusion that for the
processes with zero momentum transfer the problem should be formulated on the finite
time interval. The results are summarized and discussed in Sect. 8. The limiting
cases and effects of absorption and coherent forward scattering are discussed as well.
Section 9 contains the conclusion.

\section{Potential model}
In this section we touch briefly on the main results and the range of applicability
of potential model in the case of $n\bar{n}$ transitions. The chief drawback of this
model is given as well.

Let $U_n=$const. and $U_{\bar{n}}=$const. be the neutron potential and optical
potential of $\bar{n}$, respectively. The background field $U_n$ is included
in the unperturbed Hamiltonian $H_0$. The interaction Hamiltonian has the form
\begin{eqnarray}
{\cal H}_I={\cal H}_{n\bar{n}}+{\cal H},\nonumber\\
{\cal H}_{n\bar{n}}=\epsilon \bar{\Psi }_{\bar{n}}\Psi _n+H.c.,\nonumber\\
{\cal H}=({\rm Re}V+i{\rm Im}V)\bar{\Psi }_{\bar{n}}\Psi_{\bar{n}},\nonumber\\
V=U_{\bar{n}}-U_n={\rm Re}U_{\bar{n}}-U_n-i\Gamma /2.
\end{eqnarray}
Here ${\cal H}_{n\bar{n}}$ and ${\cal H}$ are the Hamiltonians of $n\bar{n}$
transition [11] and $\bar{n}$-medium interaction, respectively; $\epsilon $
a small parameter $\epsilon =1/\tau _{n\bar{n}}$, where $\tau _{n\bar{n}}$
is the free-space $n\bar{n}$ oscillation time.

The model can be realized by means of diagram technique [5,12] or equations of
motion [10,11,13]:
\begin{eqnarray}
(i\partial_t-H_0)n(x)=\epsilon \bar{n}(x),\nonumber\\
(i\partial_t-H_0-V)\bar{n}(x)=\epsilon n(x),\nonumber\\
H_0=-\nabla^2/2m+U_n,
\end{eqnarray}
$\bar{n}(0,{\bf x})=0$. For $V=$const. in the lowest order in $\epsilon $ the
probability of finding an $\bar{n}$ in a time $t$ is found to be
\begin{equation}
W_{\bar{n}}(t)=\frac{\epsilon ^2}{\mid\! V\!\mid ^2}[1-2\cos({\rm Re}Vt)e^
{-\Gamma t/2}+e^{-\Gamma t}].
\end{equation}
In the following we focus on the most important case $\Gamma t\gg 1$. Then
$W_{\bar{n}}(t)\sim \epsilon ^2/\mid\! V\!\mid ^2\ll 1$. The total $n\bar{n}$
transition probability $W^{{\rm pot}}_t$ (more precisely, the probability of
finding an $\bar{n}$ or annihilation products) is given by
\begin{equation}
W^{{\rm pot}}_t(t)=2\epsilon ^2t\frac{\Gamma /2}{({\rm Re}V)^2+(\Gamma /2)^2}
\approx \frac{4\epsilon ^2t}{\Gamma }.
\end{equation}
The index "pot" signifies that the non-hermitian Hamiltonian (3) is used.

The free-space $n\bar{n}$ transition probability $W_f$ is $W_f=\epsilon^2t^2$.
Comparing with Eq. (6), one obtains the suppression factor $R_{{\rm pot}}$:
\begin{equation}
R_{{\rm pot}}=\frac{W^{{\rm pot}}_t}{W_f}=\frac{\Gamma }{|V|^2t}\sim \frac{1}{|V|t}
\ll 1.
\end{equation}

The energy gap $\delta E=V$ leads to very strong suppression of $n\bar{n}$
transition in the medium and changes the functional structure of the result:
$W_f\sim t^2\rightarrow W^{{\rm pot}}_t\sim t$. Because of this, the particular
attention is given to the suppression mechanism and origin of $\delta E$.

The energy gap is the antineutron self-energy: $\delta E=V=\Sigma $. Indeed,
the result (6) can be obtained by means of diagram technique:
\begin{equation}
W_t^{{\rm pot}}=2{\rm Im}T_{ii}t,
\end{equation}
\begin{equation}
T_{ii}=-\epsilon G^{{\rm pot}}\epsilon ,
\end{equation}
$$
G^{{\rm pot}}=\frac{1}{\epsilon _n-{\bf p}_n^2/2m-U_{\bar{n}}}=-\frac{1}{V}.
$$
Here $G^{{\rm pot}}$ is the antineutron propagator, $p=(\epsilon _n,{\bf p}_n)$
is the neutron 4-momentum; $\epsilon _n={\bf p}_n^2/2m+U_n$.

Consider now the range of applicability of model (4). The total $n\bar{n}$
transition probability $W_t$ is
\begin{equation}
W_t=W_{\bar{n}}+W_a,
\end{equation}
where $W_a$ is the probability of finding the annihilation mesons ({\it i.e.} the
process (2) probability). The potential model describes correctly the $W_{\bar{n}}$.
However, for the calculation of $W_t$ and $W_a$ it is inapplicable. In the
one-particle model the total process probability $W^{{\rm pot}}_t$ is obtained by
means of Eq. (8) which follows from the unitarity condition. Since the Hamiltonian
(3) is non-hermitian (in the first approximation one can put ${\rm Re}V=0$), the
$S$-matrix is non-unitary and Eq. (8) is invalid [5]. The condition of probability
conservation
$$
1=\mid S_{ii}\mid ^2+\sum_{f\neq i}\mid T_{fi}\mid ^2
$$
can be used only if the $S$-matrix is unitary or unitarized.

It must be emphasized that in the problem under study the unitarity of $S$-matrix
is of particular importance because ${\rm Im}V$ enters the leading diagram and
plays the crucial rule (see (9)).

As a result, in the potential model the effect of final state absorption acts in
the opposite (wrong) direction [6], which is not surprising, since the unitarity
condition and non-unitary $S$-matrix are mutually incompatible. The condition
$SS^+=1$ is applied to the essentially non-unitary $S$-matrix. (The potential
model does not describe the $W_t$ at all. The non-unitarity is only formal
manifestation of this fact.) As shown below, the potential model also does not
describe the competition between scattering and annihilation of $\bar{n}$ in
the intermediate state and time-dependence of process (2). The greater the
$|{\rm Im}V|$, the greater an error in the $W_t^{{\rm pot}}$ and $W_a$. So Eq.
(6) is incorrect. The direct calculation of the antineutron absorption (process
(2)) is called for.

\section{Free-space process}
First of all we consider the free-space $\bar{n}N$ annihilation (see Fig. 1a)
and free-space process
\begin{equation}
n+N\rightarrow \bar{n}+N\rightarrow M
\end{equation}
shown in Fig. 1b.

\begin{figure}[h]
  {\includegraphics[height=.25\textheight]{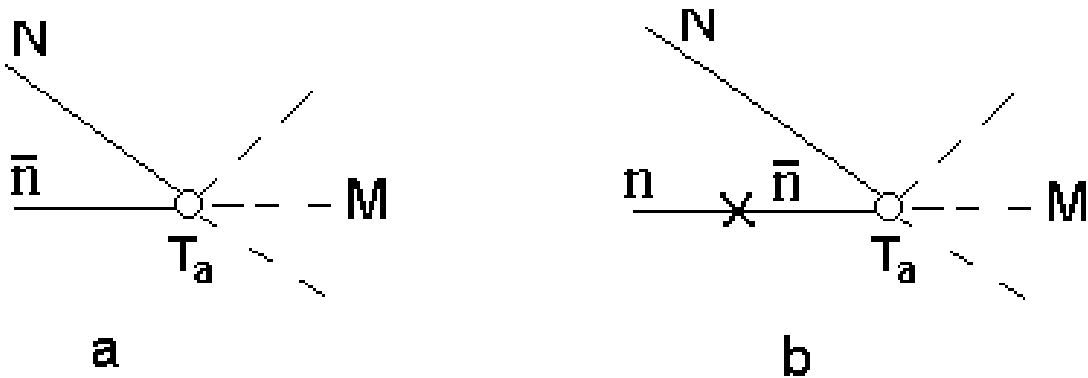}}
  \caption{(a) Free-space $\bar{n}N$ annihilation. (b) Free-space process
$n+N\rightarrow \bar{n}+N\rightarrow M$.}
\end{figure}

The matrix element of $S$-matrix $T_a$ and amplitude $M_a$ corresponding to
Fig. 1a are defined as
\begin{equation}
iT_a=<\!M\!\mid T\exp (-i\int dx{\cal H}_{\bar{n}N}(x))-1\mid\! \bar{n}N\!>=
N_a(2\pi )^4\delta ^4(p_f-p_i)M_a.
\end{equation}
Here $<\!M\!\mid $ represents the annihilation mesons, ${\cal H}_{\bar{n}N}$
is the Hamiltonian of $\bar{n}N$ interaction, $N_a$ includes the normalization
factors of the wave functions.

$T_a$ and $M_a$ involve all the $\bar{n}N$ interactions followed by
annihilation including the {\em $\bar{n}N$ rescattering in the initial state}.
The same is true for the subprocess of $\bar{n}N$ annihilation involved in Fig.
1b: the block $T_a$ should contain all the $\bar{n}N$ interactions followed by
annihilation.

We write the formulas corresponding to Fig. 1b. The interaction Hamiltonian is
\begin{equation}
{\cal H}_I={\cal H}_{n\bar{n}}+{\cal H}_{\bar{n}N}.
\end{equation}
Formally, in the lowest order in ${\cal H}_{n\bar{n}}$ the amplitude of
process (11) is given by
\begin{eqnarray}
M_{1b}=\epsilon G_0M_a,\nonumber\\
G_0=\frac{1}{\epsilon _n-{\bf p}_n^2/2m}.
\end{eqnarray}
Here $G_0$ is the antineutron propagator. Since $M_a$ contains all the
$\bar{n}N$ interactions followed by annihilation, $G_0$ is bare. We emphasize
this fact as it gives an insight into origin of $\Sigma $.

Due to the zero momentum transfer in the $n\bar{n}$-transition vertex the
4-momenta of $n$ and $\bar{n}$ are equal. The both pre- and post-$n\bar{n}$
conversion spatial wave functions of the system coincide: $\mid\! Nn_p\!>
_{{\rm sp}}=\mid\! N\bar{n}_p\!>_{{\rm sp}}$. Actually this is true for any
neutron state (for any nuclear model).

For the time being we do not go into singularity $G_0\sim 1/0$. It results
from the zero momentum transfer in the vertex corresponding to ${\cal H}_{n\bar
{n}}$. The value of $\Sigma $ is disconnected with ${\cal H}_{n\bar{n}}$ and we
want to separate these problems. The general consideration is given in Sect. 6.

\section{$n\bar{n}$ transitions in the medium}
In this section the origin of $\Sigma $ (energy gap) is studied in the framework
of microscopic theory. It is shown that the value of $\Sigma $ is uniquely
determined by the definition of the annihilation amplitude of $\bar{n}$ in the
medium. It turns out that for a realistic definition $\Sigma =0$. Also we consider
the competition between scattering and annihilation of $\bar{n}$ in the
intermediate state.

\subsection{$S$-matrix approach}
Let us consider the process (2). (The $n\bar{n}$ transitions with $\bar{n}$
in the final state are considered in the next section.) We use the scheme
identical to that for process (11) with the substitution $\bar{n}N\rightarrow
(\bar{n}-\mbox{medium})$. The background field $U_n$ is included in the neutron
wave function (Hamiltonian $H_0$); the quadratic terms ${\cal H}_{n\bar{n}}$ are
included in the ${\cal H}_I$:
\begin{eqnarray}
{\cal H}_I={\cal H}_{n\bar{n}}+{\cal H},\nonumber\\
H(t)=\int d^3x{\cal H}(x).
\end{eqnarray}
Here $H$ is the hermitian Hamiltonian of $\bar{n}$-medium interaction. The sole
physical distinction with the model (4) is in the Hamiltonian ${\cal H}$. Recall
that the potential model does not describe the processes (2) and (11) [5,6].

The process amplitude $M_2$ is uniquely determined by the Hamiltonian (15)
\begin{eqnarray}
M_2=\epsilon G_0^mM_a^m,\nonumber\\
G_0^m=\frac{1}{\epsilon _n-{\bf p}_n^2/2m-U_n}
\end{eqnarray}
(see Fig. 2a). The matrix element of $S$-matrix $T_{fi}^{\bar{n}}$ and amplitude
of antineutron annihilation in the medium $M_a^m$ are 
\begin{eqnarray}
iT_{fi}^{\bar{n}}=<\!f\!\mid T^{\bar{n}}\mid\!0\bar{n}_p\!>=
N(2\pi )^4\delta ^4(p_f-p_i)M_a^m,\nonumber\\
T^{\bar{n}}=T\exp (-i\int_{-\infty}^{\infty}dtH(t))-1
\end{eqnarray}
(compare with Eq. (12)). Here $\mid\! 0\bar{n}_p\!>$ is the state of the medium
containing the $\bar{n}$ with the 4-momentum $p$, $<\!f\!\mid $ denotes the
annihilation products, $N$ includes the normalization factors.

\begin{figure}[h]
  {\includegraphics[height=.25\textheight]{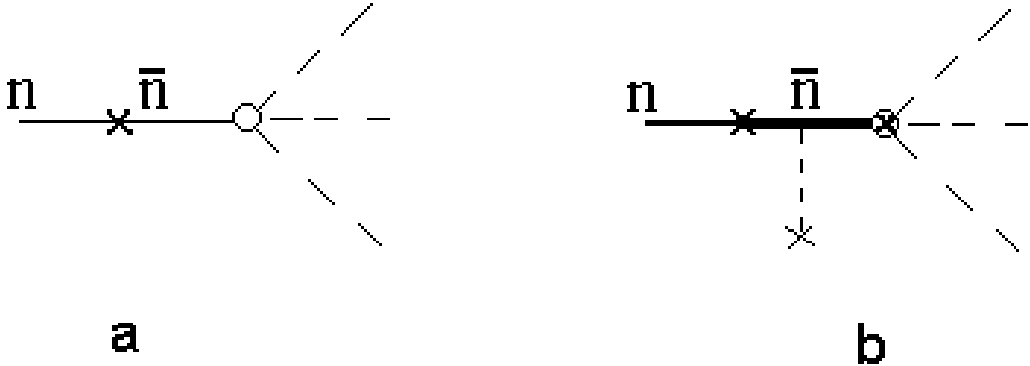}}
  \caption{(a) $n\bar{n}$ transition in the medium followed by annihilation.
The antineutron annihilation is shown by a circle. (b) Same as (a) but the
antineutron propagator is dressed (see text).}
\end{figure}

The definition of the annihilation amplitude $M_a^m$ through Eqs. (17) is natural.
If the number of particles of medium is equal to unity, Eq. (17) goes into (12).
The annihilation width $\Gamma $ is expressed through $M_a^m$:
$\Gamma \sim \int d\Phi \mid\!M_a^m\!\mid ^2$. Since $H$ appears only
in the $M_a^m$, the antineutron propagator $G_0^m$ is bare. In the next section
we perform the rigorous calculation of $M_2$.

It is important that $M_2\sim M_a^m$. The value of $\Gamma $ and corrections to
$M_a^m$ (if they are possible) have little effect on the results.

Construct now the model with the dressed propagator (see Fig. 2b). In the
Hamiltonian ${\cal H}$ we separate out the real potential $V={\rm Re}
U_{\bar{n}}-U_n$:
\begin{equation}
{\cal H}=V\bar{\Psi }_{\bar{n}}\Psi _{\bar{n}}+{\cal H}'
\end{equation}
and include it in the antineutron Green function
\begin{equation}
G^m=G_0^m+G_0^mVG_0^m+...=\frac{1}{(1/G_0^m)-V}=-\frac{1}{V}.
\end{equation}
Then
\begin{equation}
M_2=\epsilon G^mM'_a,
\end{equation}
\begin{equation}
G^mM'_a =G_0^mM_a^m.
\end{equation}
The propagator $G^m$ is dressed: $\Sigma =V\neq 0$. According to (21), the
expressions for the propagator and vertex function are uniquely connected
(if $H_I$ is fixed). The "amplitude" $M'_a(V,H')$ should describe the
annihilation. However, below is shown $M'_a$ and model (20) are unphysical.

We recall the amplitude $M_a$ involves all the $\bar{n}N$ interactions
followed by annihilation including rescattering in the initial state.
Similarly, $M_a^m$ involves all the $\bar{n}$-medium interactions followed by
annihilation including the antineutron rescattering in the initial state.
Compare now the left- and right-hand sides of (21).

From the physical point of view model (20) has no justification on
the following reasons.

1) If the number of particles of medium $n$ is equal to unity, the model (20)
does not describe the free-space process (11) because Eq. (14) contains the
bare propagator.

2) The observable values ($\Gamma $ for example) are expressed through $M_a^m$
and not $M'_a$. Compared to $M_a^m$, $M'_a$ is truncated because the portion
of the Hamiltonian $H$ is included in $G^m$. $M'_a$ has not a physical meaning.

(The formal expression for the dressed propagator should contain the annihilation
loops as well. In this case the statements given in pp. 1) and 2) are only
enhanced.)

3) Equations (19) and (20) mean that the annihilation is turned on upon forming
of the self-energy part $\Sigma =V$ (after multiple rescattering of $\bar{n}$).
This is counter-intuitive since at the low energies [15-17]
\begin{equation}
r=\frac{\sigma _{{\rm ann}}^{\bar{n}p}}{\sigma _{{\rm el}}^{\bar{n}p}}>2.5
\end{equation}
and inverse picture takes place: in the first act of $\bar{n}$-medium
interaction the annihilation occurs.

The realistic {\em competition} between scattering and annihilation should
be taken into account. Both scattering and annihilation vertices should
occur on equal terms in $M_a^m$ or $G^m$. According to pp. 1) and 2) the
latest possibility should be excluded. In line with the physical meaning of
$M_a^m$ and $M_2$, the amplitude (16) allows for the above-mentioned
effect (see Sect. 4.3).

The structure with dressed propagator like (20) arises naturally if $V$ and
${\cal H}'$ are the principally different terms and vertex function does not
depend on $V$. In the problem under consideration this is not the case.
This is evident from the formal expansion of the $T$-operator
\begin{equation}
T\exp (-i\int dx(V\bar{\Psi }_{\bar{n}}\Psi _{\bar{n}}+{\cal H}')).
\end{equation}

It is significant that even the non-realistic model (20) gives reinforcement
in comparison with the potential model [5]:
\begin{equation}
\frac{W_a}{W_t^{{\rm pot}}}=1+\left(\frac{\Gamma /2}{V}\right)^2>1
\end{equation}
because for the model (20) the probability of process (2) is
\begin{equation}
W_a\sim \Gamma
\end{equation}
instead of $W_t^{pot}\sim 1/\Gamma $.

To summarize, the introducing of dressed propagator $G^m$ (energy gap) into
process model entails an uncertainty of the vertex function $M'_a$. The
all-important effect of the competition is not taken into account. The
limiting case $n=1$ is not reproduced. $M'_a$ is unknown and unphysical.

We do not see the reasons for existence of field $V$ which should be
included into $G^m$ and thus the antineutron propagator is bare. For the
process sown in Fig. 8b the propagator is bare as well. Essentially, this
fact is governing. Below we assume the definition (17) and consequently
the model with the bare propagator (16).

\subsection{Simplest model}
The fact that the antineutron propagator is bare is obvious in the model
containing the annihilation vertex only. We consider Fig. 1a. Assume that
\begin{equation}
{\cal H}_{\bar{n}N}=\Phi ^*_M g_a\Psi _{\bar{n}}\Psi _N,
\end{equation}
where $\Phi _M$ denotes the fields of mesons. The diagrams of the $\bar
{n}N$ annihilation are shown in Fig. 3.

\begin{figure}[h]
  {\includegraphics[height=.25\textheight]{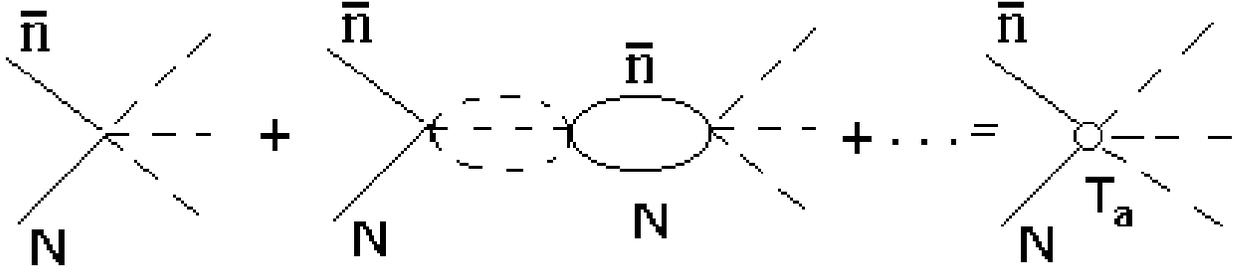}}
  \caption{$\bar{n}N$ annihilation. The interaction Hamiltonian is given by
Eq. (26).}
\end{figure}

\begin{figure}[h]
  {\includegraphics[height=.25\textheight]{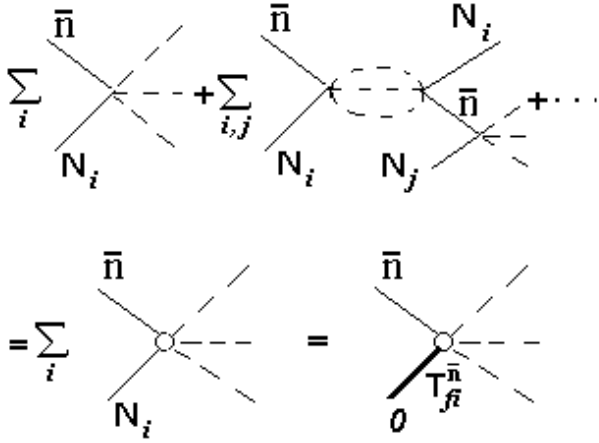}}
  \caption{Antineutron annihilation in the medium.}
\end{figure}

\begin{figure}[h]
  {\includegraphics[height=.25\textheight]{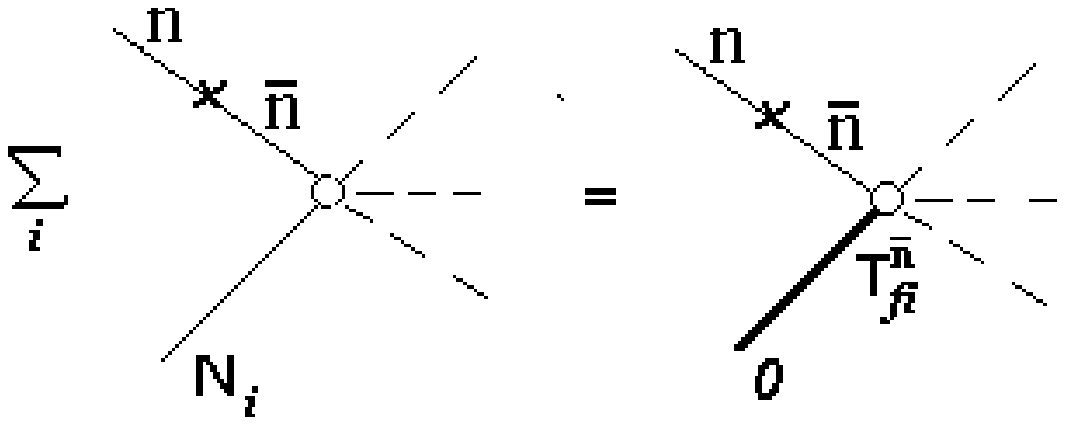}}
  \caption{$n\bar{n}$ transition in the medium followed by annihilation.}
\end{figure}

Similarly, for the $\bar{n}$-medium annihilation we take
\begin{equation}
{\cal H}={\cal H}_a=\sum_{i}\Phi ^*_M g_a\Psi _{\bar{n}}\Psi _{N_i}.
\end{equation}
The corresponding diagrams are shown in Fig. 4.

Consider now the process (2) using the same Hamiltonian ${\cal H}_a$.
The diagram is shown in Fig. 5; the Hamiltonian is given by Eqs. (15), where
${\cal H}={\cal H}_a$. The antineutron propagator is bare. The questions
connected with the self-energy part do not arise in principle because ${\cal
H}_a$ must appear only in the $T_{fi}^{\bar{n}}$ (see Fig. 4). The block
$T_{fi}^{\bar{n}}$ is described by Eqs. (17) and (27).

In view of Eq. (22), the models like (27) are reasonable and so it seems
obvious that the antineutron propagator is bare.

\subsection{Scattering and annihilation of $\bar{n}$ in the intermediate
state}
In the low-density limit the relative annihilation probability of the
intermediate antineutron $r_1$ is [15-17]
\begin{equation}
r_1=\frac{\sigma _a}{\sigma _t}>0.7,
\end{equation}
$\sigma _t=\sigma _a+\sigma _s$, where $\sigma _a$ and $\sigma _s$ are the
cross sections of free-space $\bar{n}N$ annihilation and $\bar{n}N$ scattering,
respectively. The ratio (28) or (22) is very important for the correct model
construction.

The model given above reproduces the magnitudes of $r$ and $r_1$. Indeed, let us
consider the free-space process
\begin{equation}
n+N\rightarrow \bar{n}+N\rightarrow f,
\end{equation}
where $f$ denotes $\bar{n}N$ or $M$. The annihilation and scattering channels
are defined by (11) and
\begin{equation}
n+N\rightarrow \bar{n}+N\rightarrow \bar{n}+N,
\end{equation}
respectively. The corresponding diagrams are shown in Figs. 1b and 6a. Using the
amplitude (14), the cross section of process (11) is found to be
\begin{equation}
\sigma ^{nN}_a=N\int d\Phi \mid\!M_{1b}\!\mid ^2=a^2N\int d\Phi \mid\!M_a\!\mid ^2=
a^2\sigma _a,
\end{equation}
$a=\epsilon G_0$. The normalization multiplier $N$ is the same for $\sigma ^{nN}_a$
and $\sigma _a$.

\begin{figure}[h]
  {\includegraphics[height=.25\textheight]{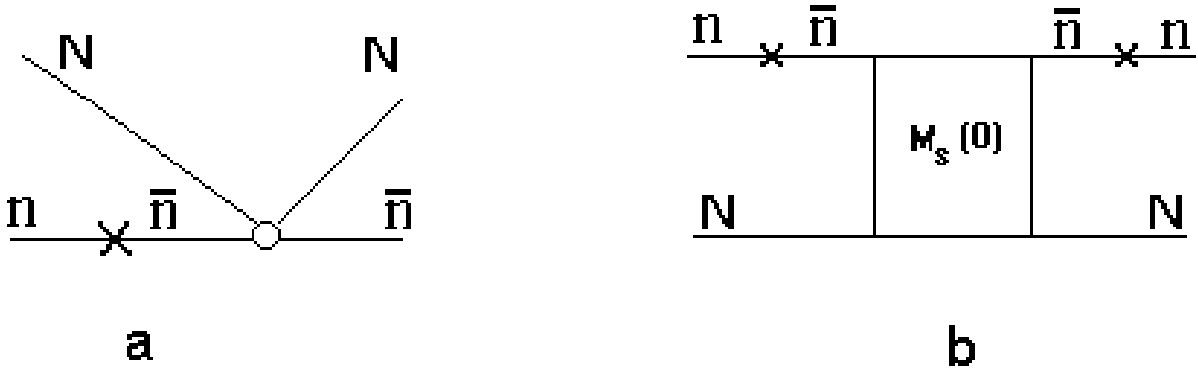}}
  \caption{Free-space processes $n+N\rightarrow \bar{n}+N\rightarrow
\bar{n}+N$ (Fig. a) and $n+N\rightarrow \bar{n}+N\rightarrow n+N$ (Fig. b).}
\end{figure}

For process (30) the similar calculation gives
\begin{equation}
\sigma ^{nN}_s=a^2\sigma _s
\end{equation}
and correspondingly
\begin{equation}
\frac{\sigma _a^{nN}}{\sigma _s^{nN}}=\frac{\sigma _a}{\sigma _s}=r.
\end{equation}
The model (13) reproduces the ratio $r$.

For $n\bar{n}$ transition in the medium $r_1$ will be calculated by means of
optical theorem. To check this calculation we obtain $r_1$ for free-space process
(29) by means of optical theorem as well. The on-diagonal matrix element (see Fig.
6b) is
\begin{equation}
M(0)=\epsilon G_0M_s(0)G_0\epsilon =a^2M_s(0),
\end{equation}
where $M_s(0)$ is the zero angle $\bar{n}N$ scattering amplitude. Let 
$\sigma ^{nN}_t$ be the total cross section of process (29). Using the optical 
theorem in the left- and right-hand sides of (34), we get
\begin{equation}
\sigma ^{nN}_t=a^2\sigma _t
\end{equation}
and
\begin{equation}
\frac{\sigma _a^{nN}}{\sigma _t^{nN}}=\frac{\sigma _a}{\sigma _t}=r_1.
\end{equation}
For process (29) the relative probability of the annihilation channel is
given by (28), as we wished prove.

In the medium instead of (11) and (29) one should consider the process (2) and
inclusive $n\bar{n}$ transition
\begin{equation}
(n-\mbox{medium})\rightarrow (\bar{n}-\mbox{medium})\rightarrow f_m,
\end{equation}
respectively. Here $f_m$ denotes $M$ or $\bar{n}$. The result is the same (see
Appendix A): for process (37) the relative annihilation probability of the
intermediate $\bar{n}$ is given by (28).

Ratio (28) is explicitly used only in the classical models like cascade one
[18]. However, the Hamiltonian should contain all the needed information, which
allows the calculation of $r$ or $r_1$. The fact that the model reproduces these
ratios is very important; otherwise one can get a wrong, {\em additional}
suppression as in model (20). Since the potential model does not describe the
processes (11) and (2), it cannot reproduce (33) and (79).

The principal results of Sect. 4 are as follows. (a) The antineutron propagator
is bare and singular. (b) In the low-density limit the ratio (28) should be
reproduced. This can be considered as a necessary condition for the correct
model construction. Model (15) satisfies this requirement.

\section{Field-theoretical approach with finite time interval}
The model must satisfy the following requirements: a) The $S$-matrix should be
unitary. b) The model should reproduce the free-space process shown in Fig. 1
and competition between scattering and annihilation considered above. These
conditions are obvious, however they are not fulfilled in the potential model.

The interaction Hamiltonian is given by Eq. (15). We use the basis $(n,\bar{n})$.
The results do not depend on the basis. A main part of existing calculations
have been done in $n-\bar{n}$ representation. The physics of the problem is in
the Hamiltonian. The transition to the basis of stationary states is a formal
step. It is possible only in the case of the potential model $H=V=$const., when
the Hamiltonian of $\bar{n}$-medium interaction is replaced by the effective
mass $H\rightarrow  m_{{\rm eff}}={\rm Re}V-i\Gamma /2$. Since the calculation
of process (2) will be done beyond the potential model, the procedure of
diagonalization of mass matrix is unrelated to our problem.

The $S$-matrix amplitudes corresponding to Figs. 1b and 2a are singular as
$G_0\sim 1/0$ and $G_0^m\sim 1/0$. Contrary to quantum electrodynamics, the
formal sum of series in $\epsilon $ gives the meaningless self-energy $\Sigma
\sim \epsilon ^2/0$. This is because the Hamiltonian ${\cal H}_{n\bar{n}}$
corresponds to 2-tail. There is no compensation mechanism by radiative
corrections.

For solving the problem the FTA is used [14]. It is infrared free. The
calculation is performed by means of the evolution operator $U(t,0)$. The 
limiting transition $t\rightarrow \infty $ is not made as it is physically 
incorrect. The FTA can be used for any problem since for the nonsingular diagrams 
it converts to the $S$-matrix approach (see Sect. 6.1).

\subsection{$n\bar{n}$ transitions with $\bar{n}$ in the final state}
First of all we consider the $n\bar{n}$ transitions with $\bar{n}$ in the
final states on the {\em finite time interval} $(t,0)$ (see Fig. 7).

\begin{figure}[h]
  {\includegraphics[height=.25\textheight]{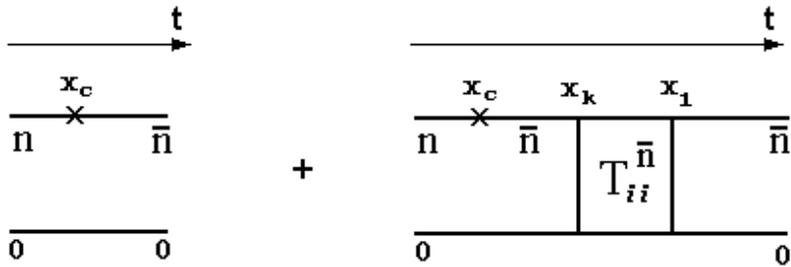}}
  \caption{$n\bar{n}$ transition in the medium with $\bar{n}$ in the final state.}
\end{figure}

We introduce the evolution operator $U(t,0)=I+iT(t,0)$. In the lowest order
in $\epsilon $ the matrix element $T_{\bar{n}n}$ is given by
\begin{equation}
<\!\bar{n}0\!\mid U(t,0)-I\mid\!0n\!>=iT_{\bar{n}n}(t,0)=-i<\!\bar{n}_p0\!\mid
\int_{0}^{t}dt_cH_{n\bar{n}}(t_c)+T^{\bar{n}}(t,0)\int_{0}^{t_k}dt_c
H_{n\bar{n}}(t_c)\mid\! 0n_p\!>,
\end{equation}
\begin{equation}
T^{\bar{n}}(t,t_c)=T\exp (-i\int_{t_c}^{t}dt_1H(t_1))-1=
\sum_{k=1}^{\infty}(-i)^k \int_{t_c}
^{t}dt_1...\int_{t_c}^{t_{k-1}}dt_kH(t_1)...H(t_k),
\end{equation}
where $\mid \!0n_p\!>$ and $\mid\!0\bar{n}_p\!>$ are the states of the medium
containing the neutron and antineutron with the 4-momentum $p=(\epsilon _n,{\bf
p}_n)$, respectively; $T^{\bar{n}}$ is the $T$-operator of $\bar{n}$-medium
interaction (compare with (17)).

We expand the $\Psi $-operators in the eigenfunctions of unperturbed Hamiltonian
$H_0=-\nabla ^2/2m+U_n$. Taking into account that $H_{n\bar{n}}\mid\!0n_p\!>=
\epsilon \mid\!0\bar{n}_p\!>$, we change the order of integration [14] and obtain
\begin{equation}
T_{\bar{n}n}(t,0)=-\epsilon t -\epsilon \int_{0}^{t}dt_ciT^{\bar{n}}_{ii}(t-t_c),
\end{equation}
$$
iT^{\bar{n}}_{ii}(\tau)=<\!\bar{n}_p0\!\mid T^{\bar{n}}(\tau) \mid\! 0
\bar{n}_p\!>,
$$
where $\tau =t-t_c$, $T^{\bar{n}}(\tau)=T^{\bar{n}}(t,t_c)$.  The
$\bar{n}$-medium interaction is separated out in the block $T^{\bar{n}}_{ii}
(\tau)$. This equation is important since the structure of matrix element
corresponding to the process (2) is similar (see (64)). On the other hand, Eq.
(40) can be verified with the use of the exactly solvable potential model.
\subsection{Verification of FTA}
To verify the FTA we obtain the results (5) and (6) of the potential model. As
in Sect. 2, we take $H=V=$const. The block $T^{\bar{n}}_{ii}(\tau)$ is easily
evaluated, resulting in
\begin{equation}
iT^{\bar{n}}_{ii}(\tau)=U^{\bar{n}}_{ii}(\tau)-1=\exp (-iV\tau)-1.
\end{equation}
The  probability of finding an $\bar{n}$ is
\begin{equation}
W_{\bar{n}}(t,0)=\mid\!T_{\bar{n}n}(t,0)\!\mid ^2.
\end{equation}
By means of Eqs. (40) and (41) it is easy to verify that $\mid\!T_{\bar{n}n}
(t,0)\!\mid ^2$ coincides with Eq. (5).

The total $n\bar{n}$ transition probability $W_t$ is given by
\begin{equation}
W_t(t,0)=1-\mid\!U_{ii}(t,0)\!\mid ^2\approx 2{\rm Im}T_{ii}(t,0),
\end{equation}
where $U_{ii}(t,0)=<\!n_p0\!\mid U(t,0)\mid\! 0n_p\!>$. In the framework of the
FTA the on-diagonal matrix element $T_{ii}$ has been calculated in Ref. [14]:
\begin{equation}
T_{ii}(t,0)=i\epsilon ^2t^2/2 -\epsilon ^2\int_{0}^{t}dt_{\alpha }\int_{0}
^{t_{\alpha }} dt_c  T^{\bar{n}}_{ii}(t_{\alpha }-t_c).
\end{equation}
Using Eqs. (41) and (44), one obtains that $2{\rm Im}T_{ii}=W_t^{{\rm pot}}$.

Consequently, the FTA reproduces all the results of the potential model. This 
was to be expected since one and the same Hamiltonian was used. The same is 
also true for any $ab$ transitions: $n\bar{n}$, $K^0\bar{K}^0$, neutrino
oscillations. (The generalization for the relativistic case is trivial.)

\subsection{Cancellation of divergences in the potential model}
The consideration given above is infrared free. Let us return to the $S$-matrix
problem formulation $(\infty ,-\infty )$. Due to the zero momentum transfer in
the $\epsilon $-vertex, any matrix element contains the singular propagator (see
Figs. 7 and 8a). However, the matrix element of potential model $T_{ii}$
obtained by means of $S$-matrix approach is not singular (see Eq. (9)).
The same is true for the process shown in Fig. 7. From microscopic theory
standpoint the reason is as follows.

In addition to the singular propagator the matrix elements mentioned above also
contain the block $T^{\bar{n}}_{ii}$ which is a sum of the zero angle rescattering
diagrams of $\bar{n}$. As a result, the self-energy part $\Sigma=V$ appears.
The corresponding mechanism of the cancellation of divergences (the forming of the
self-energy part) is illustrated by Eq. (19), where $G_0^m\sim 1/0$.

We are interesting in off-diagonal matrix elements which do not contain the sum
mentioned above ($T^{\bar{n}}_{f\neq i}$ instead of $T^{\bar{n}}_{ii}$) and hence
diverges because one singular propagator after $\epsilon $-vertex appears in
any case. (Recall that the formal sum of series in $\epsilon $ gives the
meaningless self-energy part $\Sigma \sim \epsilon ^2/0$.)

The principal result of this section is that the FTA has been verified by the
example of the exactly solvable potential model. It is involved in the block 
$iT^{\bar{n}}_{ii}(\tau)=<\!\bar{n}_p0\!\mid T^{\bar{n}}(\tau)\mid\!0\bar{n}
_p\!>$ as a special case.

\section{$n\bar{n}$ transitions followed by annihilation}
As shown above, the FTA reproduces all the potential model results. Besides, for
non-singular diagrams it converts to the $S$-matrix theory (see Sect. 6.1). We now
proceed to the main calculation.

Let us consider the process (2) in nuclear matter (see Fig. 8a). The Hamiltonians
$H_0$ and $H_I(t)$ are the same as in Sect. 4. The 4-momenta of $n$ and $\bar{n}$
coincide. The $T^{\bar{n}}$-operator involves all the $\bar{n}$-medium
interactions. In consequence of this $\Sigma =0$. In essence, we deal with 2-step
nuclear decay: dynamical $n\bar{n}$ conversion, annihilation. Its dynamical part
last only $10^{-24}$ s. The sole distinction with respect to the decay theory is
that the FTA should be used because the antineutron propagator is singular.

\begin{figure}[h]
  {\includegraphics[height=.25\textheight]{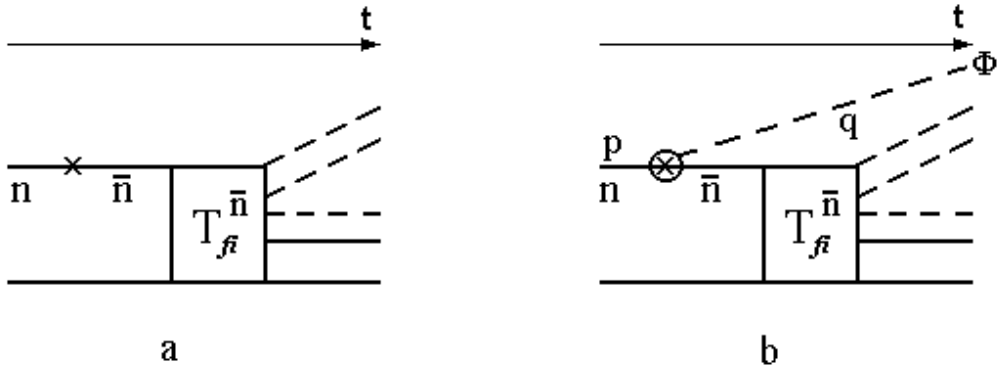}}
  \caption{(a) $n\bar{n}$ transition in the medium followed by annihilation.
(b) Same as (a) but with escaping of particle in the $n\bar{n}$
transition vertex.}
\end{figure}

We give the expressions for the amplitudes from Ref. [13]. Thereupon they will be
obtained as a special case of a more general problem. The matrix element of the
process shown in Fig. 8a is
\begin{equation}
T_{fi}(t)=-\epsilon \int_{0}^{t}dt_ciT^{\bar{n}}_{fi}(t,t_c),
\end{equation}
\begin{equation}
iT^{\bar{n}}_{fi}(t,t_c)=iT^{\bar{n}}_{fi}(\tau)=<\!f\!\mid T^{\bar{n}}(\tau)
\mid\! 0\bar{n}_p\!>.
\end{equation}
Here $T^{\bar{n}}_{fi}(\tau)$ is the matrix element of the antineutron annihilation
in a time $\tau =t-t_c$ (compare with the matrix element of $S$-matrix (17)). The  $T^{\bar{n}}(\tau)$-operator is given by (39). Similarly to (40), the
$\bar{n}$-medium annihilation is separated out in the block $T^{\bar{n}}_{fi}(\tau)$.

Consider now the more general problem. We calculate the matrix element $T_{fi}
(t)$ shown in Fig. 8b on the interval $(t/2,-t/2)$. As a result, it will be shown
that: (a) If $q\neq 0$ ($q$ is the 4-momentum of particle escaped in the $n\bar
{n}$ transition vertex) and $t\rightarrow \infty $, we come to the usual $S$-matrix
amplitude. (b) If $q\rightarrow 0$, Eq. (45) is obtained. Such scheme allows to
verify and study the FTA. Also we will see the point in which the standard
calculation scheme should be changed.

Consider the imaginary free-space decay
\begin{equation}
n\rightarrow \bar{n}+\Phi,
\end{equation}
$\Phi (x)=N_{\Phi }\exp (-iqx), N_{\Phi }=(2q_0\Omega )^{-1/2}$. For decay to
be permissible in vacuum put $m_{\bar{n}}=m-2m_{\Phi }$. As with ${\cal H}_
{n\bar{n}}$, the decay Hamiltonian ${\cal H}'_{n\bar{n}}$ is taken in the
scalar form ${\cal H}'_{n\bar{n}}=\epsilon '\bar{\Psi }_{\bar{n}}\Phi ^*
\Psi _n+H.c.$.

The corresponding process in nuclear matter is shown in Fig. 8b. This is a
nearest analogy to the process under study. The neutron wave function is
$n_p(x)=N_n\exp (-ipx)$, where $N_n=\Omega ^{-1/2}$, $p=(p_0,{\bf p})$, $p_0=
m+{\bf p}^2/2m$. The background nuclear matter field $U_n$ is omitted.

Instead of Eq. (15) we have
\begin{equation}
H_I=H'_{n\bar{n}}+H,
\end{equation}
$$
H'_{n\bar{n}}(t)=\epsilon '\int d^3x(\bar{\Psi }_{\bar{n}}\Phi ^* \Psi _n+
H.c.);
$$
$\epsilon '$ is dimensionless. In the lowest order in $H'_{n\bar{n}}$ the
matrix element $T_{fi}(t)$ is
$$
T_{fi}(t)=-<\Phi _qf0\mid \sum_{k=1}^{\infty}T_k(t)\int_{-t/2}^{t_k}dt_cH'
_{n\bar{n}}(t_c)\mid \!0n_p\!>,
$$
\begin{equation}
T_k(t)=(-i)^k \int_{-t/2}^{t/2}dt_1...\int_{-t/2}^{t_{k-1}}dt_kH(t_1)...H(t_k).
\end{equation}
Here  $<\!f\!\mid$ represents the annihilation products with $(n)$ mesons. For the
3-tail ${\cal H}'_{n\bar{n}}$ the relation $H_{n\bar{n}}\mid\!0n_p\!>=\epsilon \mid\!0\bar{n}_p\!>$ used in Sect. 5.1, is invalid. The direct calculation is
needed.

Using the standard rules of quantum field theory, we obtain (see Appendix B)
\begin{equation}
T_{fi}(t)=-i\epsilon 'N_nN_{\Phi }<f0\mid \sum_{k=1}^{\infty}(-i)T_{k-1}(t)
\int_{-t/2}^{t_{k-1}}dt_k\int d^3x_k{\cal H}'(x_k)e^{i({\bf p}-{\bf q})
{\bf x}_k}I(t_k)\mid 0\!>,
\end{equation}
\begin{equation}
I(t_k)=\int_{-\infty }^{\infty }\frac{dk_0}{2\pi }\int_{-t/2}^{t_k}dt_c
\frac{1}{k_0-m_{\bar{n}}-({\bf p}-{\bf q})^2/2m_{\bar{n}}+i0}e^{-ik_0t_k}
e^{it_c(q_0-p_0+k_0)}.
\end{equation}
From this point the calculations for Figs. 8a and 8b are essentially different.
In Eq. (51) we put $t_k=\infty $ and $-t/2=-\infty $. Then
\begin{equation}
I(\infty )=\int  \frac{dk_0}{2\pi }
\frac{e^{-ik_0t_k}}{k_0-m_{\bar{n}}-({\bf p}-
{\bf q})^2/2m_{\bar{n}}+i0} \int_{-\infty }^{\infty }dt_ce^{it_c(q_0-p_0+k_0)}
\end{equation}
and correspondingly
\begin{equation}
I(\infty )=Ge^{-i(p_0-q_0)t_k},
\end{equation}
\begin{equation}
G=\frac{1}{p_0-q_0-m_{\bar{n}}-({\bf p}-{\bf q})^2/2m_{\bar{n}}+i0},
\end{equation}
where $G$ is the non-relativistic antineutron propagator.

Let $q=(0,0)$ and $m_{\bar{n}}=m$ (see Fig. 8a). Now
\begin{equation}
G=\frac{1}{p_0-m-{\bf p}^2/2m}\sim \frac{1}{0}
\end{equation}
and $T_{fi}\sim 1/0$. This is an unremovable peculiarity. We deal with infrared
divergence, what is obvious from Fig. 8a. We thus see the specific point (the
limiting transition $t\rightarrow \infty $ in (50)) in which the standard
$S$-matrix scheme should be changed.

\subsection{Non-singular diagram}
Obtain now the amplitudes corresponding to Figs. 8a and 8b starting from (50).
If $q\neq 0$, the limit $t\rightarrow \infty $ can be considered. In Eq. (50)
we put $t\rightarrow \infty $ and substitute Eq. (53). Taking into account that
\begin{equation}
N_{\bar{n}}e^{-i(p-q)x_k}\mid \!0>=\Psi _{\bar{n}}(x_k)\mid \!\bar{n}
_{p-q}\!>
\end{equation}
and using the relation $\int d^3x_k{\cal H}'(x_k)\Psi _{\bar{n}}(x_k)=H(t_k)$
(see Appendix B), one obtains
\begin{equation}
T_{fi}=-iN_{\Phi }\epsilon 'G <f0\mid T^{\bar{n}}\mid \!0\bar{n}_{p-q}\!>.
\end{equation}
Here $\mid\!0\bar{n}_{p-q}\!>$ is the state of the medium containing the
$\bar{n}$ with the 4-momentum $p-q$, the $T^{\bar{n}}$-operator is given by (17).
With the help of the relation
$$
iT_{fi}=<\!f\!\mid T\mid\!i\!>=N(2\pi )^4\delta ^4(p_f-p_i)M_{fi}
$$
we rewrite (57) in terms of the amplitudes
\begin{equation}
M_{8b}=\epsilon 'GM^m_a.
\end{equation}
Here $M_{8b}$ is the amplitude of the process shown in Fig. 8b, $M^m_a$ is the
annihilation amplitude of $\bar{n}$ with the 4-momentum $p-q$, $G$ is given
by (54). We have obtained the usual $S$-matrix amplitude, which is the
verification of (50). As in (16), the antineutron propagator is bare.

It is easy to estimate the widths corresponding to Fig. 8b and free-space
decay (47):
$$
\Gamma _{8b}\approx \epsilon'^2\Gamma /(2\pi ^2),
$$
\begin{equation}
\Gamma_{{\rm free}}\approx \epsilon'^2m_{\Phi }/(2\pi ),
\end{equation}
where we have put $m_{\Phi }/m\ll 1$. The $t$-dependence is determined by the
exponential decay law
\begin{equation}
W_{8b,{\rm free}}=1-e^{-\Gamma _{8b,{\rm free}}t}\sim \Gamma _{8b,{\rm free}}t.
\end{equation}
These formulas will be needed below.

\subsection{Singular diagram}
Let $q=0$ and $m_{\bar{n}}=m$ (see Fig. 8a). In (50) one should put
$\epsilon '=\epsilon $ and $N_{\Phi }=1$. Upon integration with respect to
$k_0$, Eq. (51) becomes
\begin{equation}
I(t_k)=-ie^{-it_kp_0}\int_{-t/2}^{t_k}dt_c.
\end{equation}
As in (56), $N_{\bar{n}}\exp(-ipx_k)\mid \!0>=\Psi _{\bar{n}}(x_k)\mid \!
\bar{n}_p\!>$. Turning back to the Hamiltonian $H(t_k)$, one obtains
\begin{equation}
T_{fi}(t)=-i\epsilon <f0\mid \sum_{k=1}^{\infty}T_k(t)\int_{-t/2}^{t_k}dt_c
\mid \!0\bar{n}_p\!>.
\end{equation}
Using the formula
\begin{equation}
\int_{-t/2}^{t/2}dt_1...\int_{-t/2}^{t_{k-1}}dt_k\int_{-t/2}^{t_k}dt_c
f(t_1,...,t_c)=\int_{-t/2}^{t/2}dt_c\int_{t_c}^{t/2}dt_1...\int_{t_c}^{t_{k-1}}
dt_kf(t_1,...,t_c),
\end{equation}
we change the integration order and pass on to the interval $(t,0)$. Finally
\begin{equation}
T_{fi}(t)=-\epsilon \int_{0}^{t}dt_c<\!f0\!\mid T^{\bar{n}}(t-t_c)\mid\!
0\bar{n}_p\!>,
\end{equation}
$$
<\!f0\!\mid T^{\bar{n}}(\tau )\mid\!0\bar{n}_p\!>=iT^{\bar{n}}_{fi}(\tau),
$$
which coincides with (45). The result is expressed through the submatrix $T^
{\bar{n}}_{fi}(\tau )$. (Compare with (40).) Note that $T_{fi}(t)$ coincides with
the second term of (40) with the replacement $<\!i\!\mid=<\!\bar{n}_p0\!\mid
\rightarrow <\!f\!\mid $. This can be considered as a test for the $T_{fi}(t)$.

Comparing Eqs. (64) and (57), one can see the formal correspondence: if $q
\rightarrow 0$, $GT_{fi}^{\bar{n}}\rightarrow i\int_{0}^{t}d\tau T^{\bar{n}}
_{fi}(\tau )$.

\section{Infrared singularities and the formulation of the S-matrix problem}
In this section we consider the time-dependence of matrix elements and other
characteristic features of the FTA and complete the calculation of process (2)
(see also [13]).

The FTA is infrared-free. It naturally connected with the conditions of
experiment. Indeed, measurement of any process corresponds to some interval
$\tau $. So it is necessary to calculate $U_{fi}(\tau )$. The replacement
$U(\tau )\rightarrow S(\infty )$ is justified if the main contribution gives
some region $\Delta <\tau $, so that $U_{fi}(\tau >\Delta )=U_{fi}(\infty )=
S_{fi}=$const. The expressions of this type are the basis for all $S$-matrix
calculations. The following cases are possible.

1. There is bound to be asymptotic regime. Then the usual scheme realized
in field theory or non-stationary theory of scattering takes place. Fig. 8b
corresponds to this case.

2. There is no asymptotic regime. An example is provided by oscillation
hamiltonian ${\cal H}_{n\bar{n}}$. We have usual non-stationary problem. The
$S$-matrix approach is inapplicable. Because of this, for Fig. 8a the
calculation has been done in the framework of FTA.

A somewhat different explanation of application of the FTA is as follows. If
${\cal H}_I={\cal H}_{n\bar{n}}$, the solution is periodic. It is obtained by
means of non-stationary equations of motion and not $S$-matrix theory. This is
clear from the $S$-matrix definition. To reproduce the limiting case ${\cal H}
\rightarrow 0$, {\it i.e.} the periodic solution, we have to use the FTA.

Let us return to Eq. (64). The annihilation of $\bar{n}$ in nuclear matter can
be considered as the decay of a one-particle state with the characteristic time
$\tau _a$. Correspondingly, $T^{\bar{n}}_{fi}$ can be interpreted as the decay
matrix of the $\bar{n}$-medium state. Obviously
\begin{equation}
T^{\bar{n}}_{fi}(\tau >\tau _a)\approx T^{\bar{n}}_{fi}={\rm const.}
\end{equation}
and
\begin{equation}
W^{\bar{n}}=\sum_{f\neq i}\mid T_{fi}^{\bar{n}}\mid ^2=1,
\end{equation}
where $W^{\bar{n}}$ is the total decay probability of the $\bar{n}$-nucleus.
Let
\begin{equation}
t\gg \Delta \approx \tau _a.
\end{equation}
In view of this condition the submatrix $T_{fi}^{\bar{n}}$ can be calculated by
means of $S$-matrix theory. The FTA is needed only for description of the
subprocess of the $n\bar{n}$ conversion. However, the condition (66) greatly 
simplifies the calculation. One can write immediately [13]
\begin{equation}
W_a(t)\approx \sum_{f\neq i}\mid -i\epsilon tT_{fi}^{\bar{n}}\mid ^2
=\epsilon^2t^2W^{\bar{n}}=\epsilon^2t^2=W_f(t),
\end{equation}
where $W_a(t)$ is the probability of process (2).

For $n\bar{n}$ transitions in nuclear $W_t(t)=W_a(t)$ since all the $\bar{n}$
annihilate. The interpretation of $W_a(t)$ has been given above: momentary
$n\bar{n}$ conversion at some point in time between $0$ and $t$; annihilation
in a time $\tau_a\sim 6\times 10^{-24}$ s. The explanation of the
$t^2$-dependence is simple. The process shown in Fig. 8a represents two
consecutive subprocesses. The speed and probability of the whole process are
defined by those of slower subprocess. Since $\tau _a\ll t$, the annihilation
can be considered instantaneous: for any $t_1<t$ the annihilation probability
is $W^{\bar{n}}(t-t_1)\approx 1$. So, the probability of process (2) is defined 
by the speed of $n\bar{n}$ transition: $W_a\approx W_f\sim t^2$, but not $\sim 
t/\Gamma $ (see Eq. (6)). In essence, we deal with the limiting case $\tau /t
\rightarrow 0$ or, similarly, $T^{\bar{n}} _{fi}(\tau )=T^{\bar{n}}_{fi}$ at 
any $\tau $. Formally, the quadratic time-dependence follows from (64).

Owing to annihilation channel, $W_a$ is practically equal to the free-space
$n\bar{n}$ transition probability. So $\tau_{n\bar{n}}\sim T_{n\bar{n}}$,
where $T_{n\bar{n}}$ is the oscillation time of neutron bound in a nucleus.

All the results have been obtained by means of formal expansions. They are
valid at any finite $t$. Consequently, the singularities of the $S$-matrix
amplitudes $M_{1b}$ and $M_2$ result from the erroneus problem formulation. The 
problem should be formulated on the finite interval $(t,0)$. If $t\rightarrow 
\infty $, Eq. (68) diverges just as the modulus (16) squared does. The infrared 
singularities point to the fact that there is no asymptotic regime.

\section{Summary and discussion}
The importance of unitarity condition is well known [19-21]. Nevertheless, the
non-hermitian models are frequently used because on the one hand, they greatly
simplify the calculation and on the other hand, it is hoped that an error may
be inessential. This paper demonstrates that the non-unitarity of $S$-matrix
can produce a qualitative error in the results. Certainly, the unitarity is a
necessary and not sufficient condition. We compare our results and potential
model one.

The time-dependence is a more important characteristic of any process. It is
common knowledge that $t$-dependence of decay probability in the vacuum and
medium is identical. Equations (60) illustrate this fact. In our calculation
the $t$-dependencies coincide as well: $W_a\sim W_t\sim t^2$ and $W_f\sim t^2$.
The potential model gives $W_t^{{\rm pot}}\sim t$, whereas $W_f\sim t^2$. It is
beyond reason to such fundamental change.

The $\Gamma $-dependence of the results differs fundamentally as well. The
probability of the decay shown in Fig. 8b is linear in $\Gamma $
\begin{equation}
W_{8b}=\Gamma _{8b}t\sim \Gamma t
\end{equation}
(see (59) and (60)). For Fig. 8a the annihilation effect acts in the same
direction
\begin{equation}
W_a\sim W^{\bar{n}}\sim \Gamma .
\end{equation}
In the potential model the effect of absorption acts in the opposite direction
$W_t^{{\rm pot}}\sim 1/\Gamma $. Recall that the annihilation is the basic effect
determining the process speed (see (6) and (68)).

Let us consider the suppression factor $R$. From Eq. (68) we have
\begin{equation}
R=\frac{W_a}{W_f}\sim 1.
\end{equation}
For similar processes the value $R\sim 1$ is typical. Indeed, in the medium the
free-space decay (47) suppressed by the factor
\begin{equation}
\frac{\Gamma _{8b}}{\Gamma _{{\rm free}}}=\frac{\Gamma }{\pi m_{\Phi }}\approx
\frac{1}{\pi },
\end{equation}
where we have put $m_{\Phi }\approx \Gamma $.

The realistic example is the pion production $pn\rightarrow pp\pi ^-$ in vacuum
and on neutron bound in a nucleus. If the pion energy is in the region of
resonance, the pion absorption is very strong. This effects on the number of 
pions emitted from the nucleus, but not on the fact of pion formation inside the 
nucleus. (In the latter case the pion and products of pion absorption should be 
detected).

In the processes cited above $R\sim 1$. The potential model gives $R_{{\rm pot}}
\rightarrow 0$: if $\Gamma \sim 100$ MeV and $t\sim 1$ yr [22], $R_{{\rm pot}}
\sim 10^{-30}$.

Consequently, in the potential model the $t$- and $\Gamma $-dependencies are
principally incorrect. As a result, the suppression is enormous: $R_{{\rm pot}}
\rightarrow 0$. This is not surprising since the potential model describes only
$W_{\bar{n}}$. Recall that in the strong absorption region $W_{\bar{n}}\ll W_a$.

The next important point is the competition between scattering and annihilation
in the intermediate state. The models (13) and (15) reproduce the values of $r$
and $r_1$ (see Sect. 4.3). Since the potential model does not describe the
processes (2) and (11), it makes no sense to speak about competition effect in
this model. The greater the $|{\rm Im}V|$, the greater an error in the $W_t^
{{\rm pot}}$ and $W_a$ calculated by means of potential model.

Consider now the effects of coherent forward scattering and absorption. Let there
is a forward scattering alone: $H={\rm Re}V$. Since the FTA reproduces all the
potential model results (see Sect. 5.2), it describes the above-mentioned special
case as well, in particular, the suppression of oscillations by ${\rm Re}V$.

Let there is an annihilation vertex only: $V=0$ and
\begin{equation}
{\cal H}={\cal H}_a.
\end{equation}
The annihilation Hamiltonian ${\cal H}_a$ is given by (27). In this case we
inevitably arrive at the amplitude with singular propagator. The dressed propagator
cannot arise in principal (see Sect. 4.2). In view of Eq. (22) the model (73)
is {\em reasonable} and so the result $W_a\approx W_f$ seems quite natural for us.
In our calculation the approximation (73) has been not used. Nevertheless, the
result is the same as in model (73). In this connection we briefly outline
the principal points of our calculation.

The process shown in Fig. 8b is described by the Hamiltonian $H_I=H'_{n\bar{n}}
+H$. Since $H$ appears in the block $T^{\bar{n}}_{fi}$ only, the antineutron
propagator is bare. For Fig. 8a the picture is the same, however $T_{fi}\sim 1/0$
(here we keep in mind the $S$-matrix problem formulation). Due to of this we had
to use the FTA.

The fact that antineutron propagator is bare is principal. It entails the
divergence of the $S$-matrix amplitude; the application of FTA; the linear
time-dependence of the matrix element $T_{fi}(t)$ and $t^2$-dependence of the
result. In our opinion the models with dressed (and consequently non-singular)
propagator are non-realistic (see Sect. 4).

(Recall that in the potential model the antineutron propagator is dressed
as $\Sigma =V$ by the model construction. Since this model is inapplicable,
the field-theoretical approach is used. The self-energy should be considered
in the context of the concrete problem. Obviously, for Fig. 8b the propagator
is bare. For Fig. 8a it is bare as well because the $\bar{n}$-medium interaction
is the {\em same}.)

All the formulas up to (64) are true for any $ab$ transitions in which $m_a=m_b$.
(A generalizations for the relativistic case and the case when $m_a\neq m_b$ are
simple.) The next important point is the condition (67). For $n\bar{n}$
transitions in nuclei it is obvious because in this case the value $t=T_0=
1.3$ yr [22] is used ($T_0$ is the observation time in a proton-decay type
experiment). The condition $t\gg \tau _a$ leads to Eqs. (65) and (66). Due to of
them the result does not depend on a specific form of $H$ and coincides with the
result given by the model (73).

Once the antineutron annihilation amplitude is defined by (46), the rest of
calculation is rather formal. The distinguishing features of the model is
that the process amplitude is "propotional" to the annihilation amplitude $T_
{fi}\sim T^{\bar{n}}_{fi}$. This structure is typical for the direct processes.

If the condition $t\gg \Delta $ is not fulfilled, the direct calculation of (64)
is needed. However, the qualitative picture remains the same: the process
amplitude is proportional to the absorption amplitude.

It is interesting to study the behavior of $W_t$ in the intermediate range
$t\sim \Delta $. It seems plausible that $W_t$ depends slightly on the
value of $\Delta /t$ (in comparison with potential model results). We also note
that there is no asymptotic regime for free-space $K^0\bar{K}^0$ oscillations.
In our opinion, it makes sense to look at the calculation of $\Delta m=m_L-m_S$
(GIM mechanism) from the standpoint of applicability of $S$-matrix approach in
this case (see Sect. 7).

\section{Conclusion}
The approach considered above reproduces all the results on the particle 
oscillations (Sect. 5.2). Certainly, for the problems where the absorption is 
inessential, the standard model of oscillations is more handy since it is more 
simple. Our approach is oriented to the processes like (1) which are not 
described by the potential model.

The direct calculation of $n\bar{n}$ transitions in nuclear matter followed by
annihilation has been done. The results have been discussed in Sect. (8). We
confirm our restriction [13] on the free-space $n\bar{n}$ oscillation time $\tau _{n\bar{n}}>10^{16}$ yr. Compared to [13], the result (68) was obtained as a special
case of a more general problem. Besides, the medium corrections, the uncertainties
related to amplitudes and competition between scattering and annihilation in the
intermediate state have been studied. The model (73) and analysis made
in Sect. 4 show that $\Sigma =0$. Nevertheless, this is a point of great nicety.
The further investigations are desirable. The region $t<\Delta $ and oscillations
of another particles can be considered as well.

The calculation up to (64) is formal. With the replacement $T^{\bar{n}}_{fi}
(\tau )\rightarrow T^b_{fi}(\tau )$, where $T^b_{fi}$ is the $b$-particle
absorption amplitude, the matrix element (64) describes the process (1) in
which $m_a=m_b$. In this connection we point out some features of Eq. (64).
a) The amplitude $T_{fi}(t)$ is "proportional" to the amplitude $T_{fi}^b(\tau
)$. In the potential model the effect of $b$-particle absorption acts in the
opposite direction, which tends to suppress the process. b) In the
lowest order in $\epsilon $ the potential model gives the linear $t$-dependence
$W_t^{{\rm pot}}\sim t/\Gamma $. For any block $T_{fi}^b(\tau )$ model the
time-dependence of the value $\mid T_{fi}(t)\mid ^2$ cannot be linear.

Also we would like to emphasize that for a processes with zero momentum transfer
the problem should be formulated on the finite time interval.\\
\\
{\bf Appendix A}\\
In this appendix the relative annihilation probability of the intermediate
$\bar{n}$ for $n\bar{n}$ transition in the medium is calculated. Similarly to (31),
we obtain the probability of process (2) in a unit of time
\begin{equation}
\Gamma _2=N_1\int d\Phi \mid\!M_2\!\mid ^2=a^2_mN_1\int d\Phi \mid\!M_a^m\!\mid
^2=a^2_m\Gamma ,
\end{equation}
$a_m=\epsilon G_0^m$. The normalization multiplier $N_1$ is the same for $\Gamma _2$
and $\Gamma $. The term "width" is unused because the $t$-dependence of process (2)
does not need to be $\exp (-\Gamma _2t)$ (see Sect. 7).

In the low-density approximation [23,24] $\Gamma =v\rho \sigma _a$ and
\begin{equation}
\Gamma _2=a_m^2v\rho \sigma _a.
\end{equation}
The on-diagonal matrix element $M^m(0)$ corresponding to the process 
$(n-\mbox{medium})\rightarrow (\bar{n}-\mbox{medium})\rightarrow (n-\mbox{medium})$
is
\begin{equation}
M^m(0)=\epsilon G_0^mM^m_s(0)G_0^m\epsilon =a^2_mM^m_s(0)
\end{equation}
(compare with (34)). Here $M^m_s(0)$ is the amplitude of zero angle
scattering of $\bar{n}$ in the {\em medium}.

Taking into account that
\begin{equation}
\frac{1}{T_0}2{\rm Im}M^m(0)=\Gamma _t,
\end{equation}
$$
\frac{1}{T_0}2{\rm Im}M^m_s(0)=v\rho \sigma _t
$$
($T_0$ is the normalization time, $\Gamma _t$ is the probability of the
process (37) in a unit of time), one obtains
\begin{equation}
\Gamma _t=a_m^2v\rho \sigma _t
\end{equation}
and correspondingly
\begin{equation}
\frac{\Gamma _2}{\Gamma _t}=r_1.
\end{equation}
Equations (75) and (78) are interpreted in line with the low-density approximation
physics.\\
\\
{\bf Appendix B}\\
The calculation is standard [25,26] up to the integration over $t$. The
neutron and antineutron are assumed spinless. We have
\begin{equation}
\Psi _n(x)\mid \!n_p\!>=\Psi _n(x)b^+(p)\mid \!0\!>=N_ne^{-ipx}\mid \!0>,
\end{equation}
\begin{equation}
<\Phi _q\mid \Phi ^*(x)=<0\mid N_{\Phi }e^{iqx}.
\end{equation}
Then
\begin{equation}
<\Phi _q\mid \Phi ^*(x_c)\Psi _{\bar{n}}(x_c)\mid \!n_p\!>=N_nN_{\Phi }e^{i(q-p)x_c}.
\end{equation}
In the last multiplier of Eq. (49) we separate out the antineutron field operator
$\Psi _{\bar{n}}(x_k)$:
\begin{equation}
H(t_k)=\int d^3x_k{\cal H}(x_k)=\int d^3x_k{\cal H}'(x_k)\Psi _{\bar{n}}(x_k).
\end{equation}
Equation (49) becomes
\begin{equation}
T_{fi}(t)=-i\epsilon 'N_nN_{\Phi }<f0\mid \sum_{k=1}^{\infty}(-i)T_{k-1}(t)
\int_{-t/2}^{t_{k-1}}dt_k\int d^3x_k{\cal H}'(x_k)J(t_k)\mid \!0>,
\end{equation}
\begin{equation}
J(t_k)=\int_{-t/2}^{t_k}dt_c\int d^3x_c<T(\Psi _{\bar{n}}(x_k)
\bar{\Psi }_{\bar{n}}(x_c))>e^{i(q-p)x_c}.
\end{equation}
For Fig. 8a the problem is non-relativistic and so for Fig. 8b we also take the
non-relativistic antineutron propagator
\begin{equation}
<T(\Psi _{\bar{n}}(x_k)\bar{\Psi }_{\bar{n}}(x_c))>=iG(x_k-x_c)=
i\int \frac{dk_0}{2\pi }e^{-ik_0(t_k-t_c)}
\int \frac{d^3k}{(2\pi )^3}\frac{e^{i{\bf k}({\bf x}_k-{\bf x}_c)}}
{k_0-m_{\bar{n}}-{\bf k}^2/2m_{\bar{n}}+i0}.
\end{equation}
Upon integrating over ${\bf x}_c$ and ${\bf k}$ we obtain Eqs. (50)
and (51).

\end{document}